\documentclass[12pt,preprint]{aastex}

\usepackage{emulateapj5} 
\usepackage{graphics} 

\begin{document}

\slugcomment{submitted to The Astrophysical Journal Letters}
 
\shorttitle{Molecular gas in Mrk~86}
\shortauthors{Gil de Paz et al$.$}

\title{$^{12}$CO mapping of the low-metallicity BCD galaxy Mrk~86}

\author{A. Gil de Paz\altaffilmark{1,2}, S. A. Silich\altaffilmark{3,4},
B. F. Madore\altaffilmark{1,5}, C. S\'{a}nchez
Contreras\altaffilmark{2}, J. Zamorano\altaffilmark{6}, and
J. Gallego\altaffilmark{6}}

\altaffiltext{1} {NASA/IPAC Extragalactic Database, California Institute of Technology, MS 100-22, Pasadena, CA 91125; agpaz, barry@ipac.caltech.edu}
\altaffiltext{2} {Jet Propulsion Laboratory, California Institute of Technology, MS 183-900, Pasadena, CA 91109; sanchez@eclipse.jpl.nasa.gov}
\altaffiltext{3} {Instituto Nacional de Astrof\'{\i}sica, \'{O}ptica y Electr\'{o}nica, AP 51, Luis Enrique Erro 1, Tonantzintla 72000, Puebla, Mexico; silich@inaoep.mx}
\altaffiltext{4} {Main Astronomical Observatory, National Academy of Sciences of Ukraine, 03680 Kiyv-127, Golosiiv, Ukraine}
\altaffiltext{5} {The Observatories, Carnegie Institution of Astronomy, 813 Santa Barbara Street, Pasadena, CA 91101}
\altaffiltext{6} {Departamento de Astrof\'{\i}sica, Universidad Complutense de Madrid, Av$.$ Complutense s/n. E-28040 Madrid, Spain; jaz, jgm@astrax.fis.ucm.es}

\begin{abstract}

We have mapped the $^{12}$CO $J$=1-0 and $J$=2-1 line emission in
Mrk~86, one of the most metal-deficient Blue Compact Dwarf galaxies so
far detected in $^{12}$CO. The $^{12}$CO emission is distributed in a
horseshoe-like structure that follows the locus of the most recent
star formation regions. The minimum in molecular-line emission
corresponds to the position of an older, massive nuclear
starburst. The H$_{2}$ mass of the galaxy (in the range
0.4-5$\times$10$^{7}$\,M$_{\odot}$) and its morphology have been
compared with the predictions of hydrodynamic simulations of the
evolution of the interstellar medium surrounding a nuclear
starburst. These simulations suggest that the physical conditions in
the gas swept out by the starburst could have led to the formation of
the ring of molecular gas reported here. This result provides an
attractive scenario for explaining the propagation (in a galactic
scale) of the star formation in dwarf galaxies.

\end{abstract}

\keywords{ galaxies: evolution -- galaxies: starburst -- galaxies: dwarf -- galaxies: individual(Mrk~86) -- radio lines: galaxies }

\section {Introduction}
\label{sec1}

One of the most intriguing questions about the evolution of Blue
Compact Dwarf (BCD) galaxies concerns the mechanism(s) responsible for
the triggering and propagation of their star formation activity
(Noeske et al$.$ 2000; Papaderos et al$.$ 1996). The majority of BCDs
show recent star formation distributed across a large fraction of the
optical extent of the galaxy (those BCDs classified as iE by Loose \&
Thuan 1985) in contrast to the more rare nucleated BCDs (nE type). The
formation of density waves, the most commonly suggested propagation
mechanism for the star formation in grand-design spirals (see
Englmaier \& Shlosman 2000 and references therein), is inhibited in
BCDs due to their very low total mass. Therefore, a different
mechanism must be taking place in a generalized way in these galaxies.

In the case of the BCD galaxy Mrk~86, Gil de Paz, Zamorano, \& Gallego
(2000; GZG00) proposed that the most recent star formation activity in
this object was activated by the evolution of a massive nuclear
starburst that swept out large amounts of gas at distances to
0.5-1\,kpc. This gas, due to its high column density, would have
become molecular (see Franco \& Cox 1986) probably leading to the
formation of new generations of stars.

In this Letter we report the discovery of a horseshoe-like structure
of molecular gas around the nuclear starburst in Mrk~86. Using
hydrodynamic simulations of the evolution of the interstellar medium
around such a nuclear starburst, we show that the physical conditions
in the gas swept out by the starburst can plausibly lead to the
formation of such a massive molecular-gas ring.

The analysis of the $^{12}$CO-line emission in Mrk~86
(Z=1/15-1/5\,Z$_{\odot}$; GZG00) not only contributes to an
understanding of the star formation propagation mechanisms in BCDs,
but also can shed light on the poorly known excitation conditions of
the molecular gas and the CO shielding in low-metallicity galaxies
(see Walter et al$.$ 2001 and references therein). In this sense,
Mrk~86 is, after I~Zw~36 (Young et al$.$ 1995), the lowest metallicity
BCD detected so far in CO (Sage et al$.$ 1992) and, now, the first to
be mapped in that line.

\section {Observations and Results}
\label{sec2}

We have simultaneously mapped the $^{12}$CO $J$=1-0 and the $J$=2-1
emission in Mrk~86 using the IRAM 30\,m MRT at Pico de Veleta
(Granada, Spain) from 2001 May 9 to May 17. Four different receivers
were used, two operating at 1.3\,mm and two at 3\,mm (V and H
polarizations). All the spectra were obtained in wobble switching mode
with azimuthal amplitude of $\pm$4\arcmin. The pointing accuracy of
the telescope ($\pm $2\arcsec\ r.m.s.) was checked every hour using
nearby quasars. Data calibration and reduction was performed following
the standard procedure within the IRAM-GAG software package. 


The spatial resolution of these observations is 12\arcsec\ and
22\arcsec\ for CO 2-1 and 1-0, respectively. A grid of 30 observed
points covers a diamond-shaped region fully sampled in the 1-0
transition ($\Delta$x=$\Delta$y=12\arcsec).  The average time spent
per point was 40\,min (on+off). The r.m.s$.$ (in T$_{\mathrm{mb}}$
scale) of the individual CO(1-0) [CO(2-1)] spectra ranges between
2-6\,mK [4-29\,mK] for a final spectral resolution of
5.21\,km\,s$^{-1}$ [3.26\,km\,s$^{-1}$].

\begin{figure}
\caption{{\bf a)} $^{12}$CO spectra averaged over the 30 positions
observed in Mrk~86. In the upper panel the 21\,cm H{\sc i}-line
profile (in arbitrary units) obtained by Swaters (1999) is indicated
by a {\it dashed-line}. {\bf b)} $^{12}$CO spectra for the central point
(offset 0,0) of the grid.\label{fig1}}
\plotone{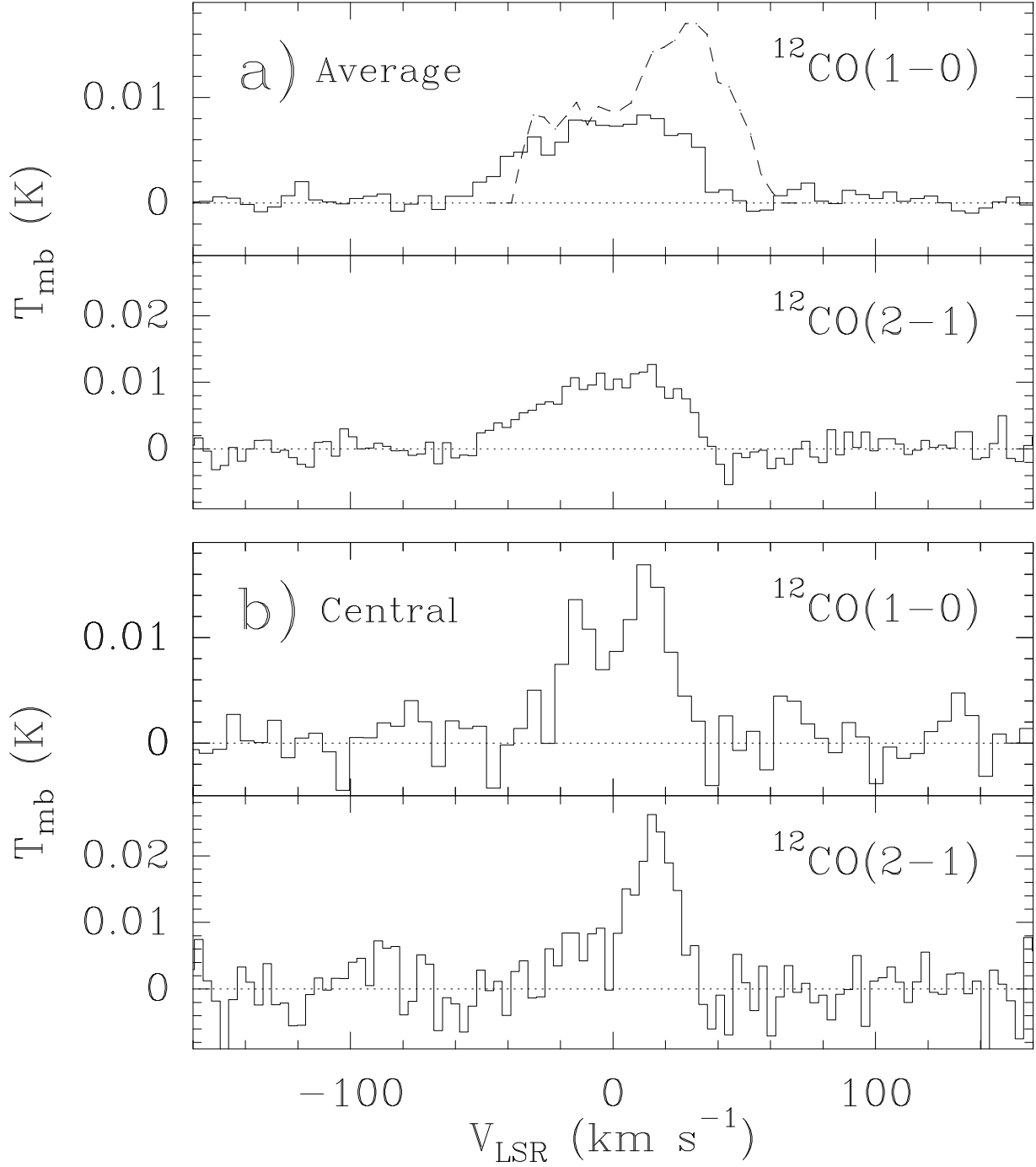}
\end{figure}

In Figure~\ref{fig1}a we show our CO spectra averaged over the 30
positions observed. For the 1-0 and 2-1 transitions the line centroids
are $v_{\mathrm{LSR}}$=442$\pm$1\,km\,s$^{-1}$ and
442$\pm$3\,km\,s$^{-1}$ and the 20\% peak line widths
89$\pm$4\,km\,s$^{-1}$ and 65$\pm$4\,km\,s$^{-1}$, respectively. The
averaged CO spectra obtained are quite symmetric around the galaxy
systemic velocity ($v_{\mathrm{sys,LSR}}$=444\,km\,s$^{-1}$). In
comparison, however, the 21\,cm H{\sc i}-line profile (Swaters 1999)
is much more intense in the interval ($v_{\mathrm{sys}}$ to
$v_{\mathrm{sys}}$$+$60\,km\,s$^{-1}$) than at lower velocities
($v_{\mathrm{sys}}$$-$60\,km\,s$^{-1}$ to $v_{\mathrm{sys}}$). The CO
spectra at the galaxy optical center are shown in
Figure~\ref{fig1}b. This position coincides with that previously
observed by Sage et al$.$ (1992) also using the IRAM 30\,m telescope.

\begin{figure}
\caption{{\bf a)} CO(1-0) integrated intensity map with the $R$-band
contours overimposed. The horseshoe-like distribution of the gas is
clearly seen around the position of the nuclear starburst. {\bf b)}
H$\alpha$ image of Mrk~86 and, overimposed, CO(2-1) integrated
intensity map (contours ranging between 1.0 and 2.0\,K\,km\,s$^{-1}$
in steps of 0.1\,K\,km\,s$^{-1}$). The north (approacing) and south
(receding) lobes of the bubble Mrk~86-B are sketched. A white
rectangle marks the orientation where the p-v diagram shown in
Figure~\ref{fig3} was extracted. The origin in both figures is the
center of the outer optical isophotes
(RA(J2000)=08$^{\mathrm{h}}$13$^{\mathrm{m}}$14$\fs$56,
DEC(J2000)=$+$45\degr 59\arcmin 30$\farcs$2; GZG00). Beam sizes are
also shown.\label{fig2}} \plottwo{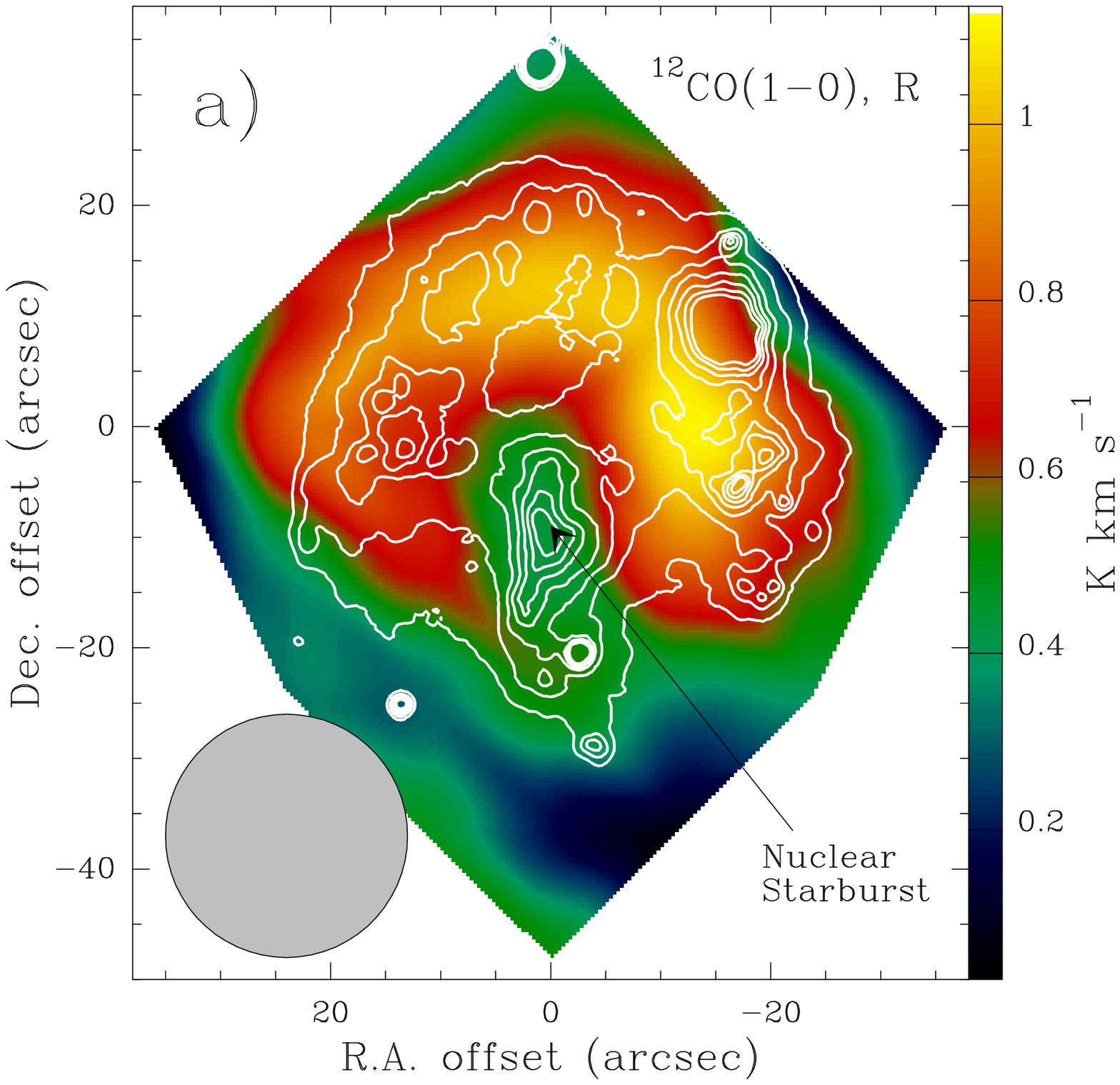}{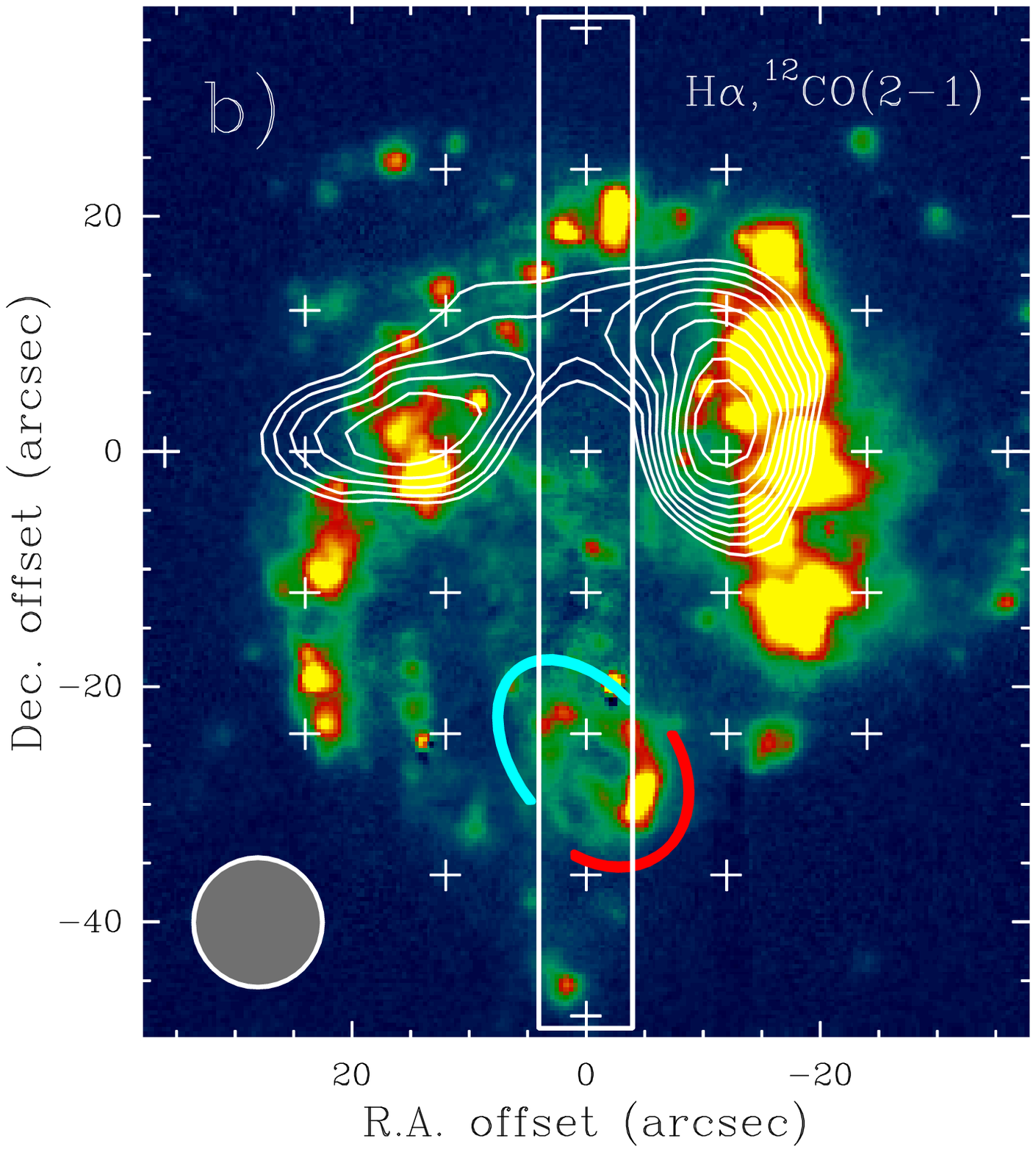}
\end{figure}

CO maps of integrated intensity are presented in
Figure~\ref{fig2}. The left panel shows the molecular gas emission
distributed in a horseshoe-like structure with a diameter of
$\sim$35\arcsec which corresponds to 1.2\,kpc (where we have assumed a
distance of 6.9\,Mpc; Sharina et al$.$ 1999; SKT99). The molecular gas
closely follows the distribution of the star-forming regions (traced
by H$\alpha$; see right panel) with a relative minimum at the position
of the nuclear starburst (GZG00). This starburst is apparent in the
R-band image 7\arcsec\ south of the galaxy center.

In Figure~\ref{fig3} we show the CO(1-0) position-velocity (p-v)
diagram along PA=0\degr. The rotation axis of Mrk~86 is oriented at
PA$\sim$80\degr\ (Gil de Paz et al$.$ 1999; GZG99 hereafter), and
therefore, the previous p-v diagram is a good representation of the
radial velocity curve of the molecular gas. The CO p-v diagram
obtained is similar to that of the ionized gas (for the same PA),
although the radial velocity gradient for the molecular gas component
seems to be slightly lower than that of the ionized gas. The
difference observed, however, is not conclusive because of the errors
in the measured velocities and, specially, the effect of beam smearing
in our CO(1-0) data. We also note in this diagram the presence of
CO(1-0) blueshifted and redshifted emission located at offset
$-$24\arcsec. These components are the molecular counterparts to the
bubble Mrk~86-B lobes detected in the optical. The bubble expansion
velocity derived from the molecular gas is 31$\pm$2\,km\,s$^{-1}$,
very similar to that obtained from the optical emission lines,
34\,km\,s$^{-1}$ (GZG99).

\section{Molecular gas physical properties}
\label{sec3}

In order to study the physical conditions in the molecular gas we have
derived the 2-1/1-0 brightness temperature ratio in Mrk~86.  The mean
values of these ratios are 1.34$\pm$0.46 and 0.40$\pm$0.14 under {\it
Uniform Filling} and {\it Point Source} approximations, respectively,
using only those positions where both transitions are clearly
detected. Although the {\it Point Source} approximation is usually
assumed for BCD galaxies (Sage et al$.$ 1992; Meier et al$.$ 2001),
our CO maps suggest that the emission in Mrk~86 arises in an extended
structure formed by several (most likely unresolved) clouds. Thus, we
have decided to use a more precise approach for deriving the 2-1/1-0
brightness temperature ratio. We deconvolve our CO(1-0) map with a
22\arcsec\ gaussian beam. Then, we determine the 1-0 and 2-1 line
intensities at each of the positions detected in both transitions
along with the size of the emitting regions (in the deconvolved map)
inside the 1-0 and 2-1 beams for each of these positions. The average
of the 2-1/1-0 brightness temperature ratios derived at each point
yields a 2-1/1-0 brightness temperature ratio of 1.06$\pm$0.40. We
have not found any systematic difference between the 2-1/1-0 ratio at
the position of the nuclear starburst and the outermost regions of the
galaxy, which indicates that there is no significant change of the
excitation conditions.

Meier et al$.$ (2001) failed to detect $^{12}$CO(3-2) at the center of
Mrk~86 and provided a very low upper limit of 0.45 to the 3-2/1-0
brigthness temperature ratio using the 1-0 intensity published by Sage
et al$.$ (1992). However, the value reported by Sage et al$.$ at this
position, 1.12\,K\,km\,s$^{-1}$, is a factor of two larger than that
measured by us, 0.56\,K\,km\,s$^{-1}$. This difference could be due to
calibration or pointing inaccuracies. Therefore, we have decided to
adopt a more conservative upper limit of 0.9 to the 3-2/1-0 ratio.

Under the LTE and optically thick assumptions we can estimate the
excitation temperature ($T_{\mathrm{ex}}$) of the molecular gas from
the 2-1/1-0 and 3-2/1-0 brightness temperature ratios. In particular,
from 2-1/1-0 we infer $T_{\mathrm{ex}}$$>$6\,K and, according to the
3-2/1-0 upper limit, $T_{\mathrm{ex}}$$<$45\,K. From the predictions
of Large Velocity Gradient models and the 2-1/1-0 and 3-2/1-0 ratios
measured we have also derived a range of compatible molecular gas
densities ($n_{\mathrm{H}_{2}}$) and kinetic temperatures
($T_{\mathrm{kin}}$). A [CO/H$_{2}$] abundance of 8$\times$10$^{-6}$
($Z_{\odot}$/10) and a velocity gradient of 1\,km\,s$^{-1}$\,pc$^{-1}$
are assumed (see e.g$.$ Meier et al$.$ 2001). The line ratios measured
suffer from a strong degeneracy in the physical properties of the
molecular gas, and only rough constraints can be obtained. We can only
conclude that, if $n_{\mathrm{H}_{2}}$$>$500\,cm$^{-3}$, then the
molecular gas cannot be warmer than $T_{\mathrm{kin}}$=40\,K.

We have also obtained a variety of estimates of the mass of molecular
gas in Mrk~86. First, we have computed the molecular mass under the
optically-thin approximation from the $^{12}$CO(1-0) integrated
luminosity ($L_{\mathrm{CO}}$ =
2$\times$10$^{6}$\,K\,km\,s$^{-1}$\,pc$^{2}$) adopting an average
value for the excitation temperature ($<$$T_{\mathrm{ex}}$$>$ =
20\,K). This yields $M_{\mathrm{thin}}$ =
4$\times$10$^{6}$\,M$_{\odot}$. It is worth noting that this
transition is rarely optically thin ($^{13}$CO and C$^{18}$O
transitions would be more appropriated) and so the value given above
provides only a rough but firm lower limit to the actual molecular
mass of the galaxy. From the $^{12}$CO(1-0) line width
($\Delta\,v_{1/2}$ = 75\,km\,s$^{-1}$) we can also estimate the
molecular gas mass assuming virial equilibrium, which leads to
$M_{\mathrm{vir}}$ = 7$\times$10$^{8}$\,M$_{\odot}$. Again, the value
derived, although represents a firm upper limit, is very far from the
actual molecular mass since the gas velocities measured are supported
by rotation. Moreover, a large fraction of the mass in the galaxy
central regions is in the form of stars (GZG99).

An estimate of the H$_{2}$ mass can be also obtained using a
$L_{\mathrm{CO}}$-to-$M_{\mathrm{H}_2}$ conversion factor appropriate
to the low metallicity of Mrk~86
($X_{\mathrm{CO}}$$\equiv$$M_{\mathrm{H}_2}$/$L_{\mathrm{CO}}$; see
e.g$.$ Barone et al$.$ 2000). The dependence of this factor on
metallicity was studied by Arimoto et al$.$ (1996; A96) using the
virial masses and CO luminosities of Giant Molecular Clouds in nearby
galaxies. These authors proposed the following relation,
log\,($X_{\mathrm{CO}}$/$X_{\mathrm{MW}}$)=$-$[O/H]$+$8.93, where
$X_{\mathrm{MW}}$ is the Milky Way conversion factor,
1.56$\times$10$^{20}$ molecules\,cm$^{-2}$\,(K\,km\,s$^{-1}$)$^{-1}$
(Hunter et al$.$ 1997). The average value that we adopt for the gas in
Mrk~86 is 1/10$^{\mathrm{th}}$ the Solar value, which yields
$X_{\mathrm{CO}}$ = 1.6$\times$10$^{21}$
molecules\,cm$^{-2}$\,(K\,km\,s$^{-1}$)$^{-1}$. From this value and
the CO luminosity given above we derive a molecular gas mass (helium
excluded) of $M_{\mathrm{H}_2}$=5$\times$10$^{7}$\,M$_{\odot}$.

Based on recent studies of low metallicity BCD galaxies we believe,
however, that the A96 relation is probably over-estimating the
molecular gas mass in those low-metallicity BCDs detected in CO, and,
in particular, the molecular mass given above. Observations of the SMC
carried out by Rubio et al$.$ (1993) have shown that the
$L_{\mathrm{CO}}$-to-$M_{\mathrm{H}_2}$ conversion factor depends on
the spatial scale of the CO emitting region considered: at spatial
scales of 10-20\,pc $X_{\mathrm{CO}}$ is only slightly higher than
$X_{\mathrm{MW}}$ while at larger scales ($\sim$100\,pc) its value
increases dramatically. This suggests the presence of large amounts of
H{\sc i} or {\it hidden} H$_{2}$ (not associated with CO) between the
dense clumps where the CO emission arises. The lack of detection of
diffuse H$_{2}$ in I~Zw~18 by FUSE (Vidal-Madjar et al$.$ 2000) and
the very low upper limit derived to the [CII]/CO line ratio in I~Zw~36
(Mochizuki \& Onaka 2001) suggest that, in the case of the
low-metallicity BCDs, most of the H$_{2}$ is probably in the form of
these dense clumps. In dwarf irregular galaxies like IC~10 the intense
[CII] emission observed has been argued as due to the presence of
large amounts of H$_{2}$ with no CO emission associated (Madden et
al$.$ 1997). These results suggest that in low-metallicity BCDs most
of the H$_{2}$ is probably associated with emitting CO and, therefore,
the global $L_{\mathrm{CO}}$-to-$M_{\mathrm{H}_2}$ conversion factor
would be similar to the local value derived from the analysis of the
CO line ratios. Noteworthly, Large Velocity Gradient modeling of the
few low-metallicity BCDs detected in $^{12}$CO favors local conversion
factors lower than the A96 predictions and similar, in some cases, to
the Galactic value (Sage et al$.$ 1992; Barone et al$.$
2000). Unfortunately the lack of observations on optically-thin
$^{13}$CO and C$^{18}$O lines, which would provide an accurate
determination of the physical conditions in the gas, makes difficult
to obtain definitive conclusions in this sense. Finally, it is also
worth noting that other authors have proposed a shallower dependence
of the $L_{\mathrm{CO}}$-to-$M_{\mathrm{H}_2}$ conversion factor on
metallicity than that argued by Arimoto et al$.$ (1996). For example,
using the relation proposed by Wilson (1995) we derive a molecular
mass of $M_{\mathrm{H}_2}$=2$\times$10$^{7}$\,M$_{\odot}$.

Therefore, we conclude that the H$_{2}$ mass of Mrk~86 is certainly in
the range 0.4-70$\times$10$^{7}$\,M$_{\odot}$ (based on the {\it
optically-thin} and {\it virial} approximation estimates) with a most
probable value in the range 0.4-5$\times$10$^{7}$\,M$_{\odot}$. The
latter range is given by the uncertainties expected in using the
metallicity-scaled $L_{\mathrm{CO}}$-to-$M_{\mathrm{H}_2}$ conversion
factor.

\section{The origin of the molecular horseshoe: Hydrodynamic simulations}

The spatial distribution of the recent star formation regions, and the
age difference between these regions (5-10\,Myr old) and the nuclear
starburst ($\sim$30\,Myr old) suggest that the development of the
starburst could have induced the recent star formation activity in
Mrk~86 (GZG00). Indeed, the energy injected during the evolution of
the nuclear starburst could have resulted in the sweeping out of large
amounts of interstellar gas to large galactocentric distances. This
gas would have eventually reached gas surface densities high enough to
lead to the formation of molecular gas and the subsequent formation of
new stars. Our CO 1-0 and 2-1 maps show that the molecular gas in this
galaxy is distributed in a horseshoe-like structure around the center
of the galaxy, which is in nice agreement with the predictions of this
scenario.

In order to check the scenario described above we have modeled the
evolution of the ISM surrounding the nuclear starburst in Mrk~86 and
compared the results obtained with the molecular-gas content and
distribution derived from our CO data. For the calculations we have
used our 3D code based on the thin layer approximation (Silich \&
Tenorio-Tagle 1998) which was modified to take into account a fast
transition from the energy-dominated to a momentum-dominated
phase. The comparison of our spherically symmetric test runs with the
analytic solution of Koo \& McKee (1992) yielded agreement to better
than 3\%.

\begin{figure}
\caption{CO(1-0) position-velocity diagram along PA=0\degr. Triangles
represent the radial velocities measured using optical emission lines,
white ones for the galaxy global velocity field (slit \#4r in GZG99)
and black ones for the bubble Mrk~86-B (slit \#4b in
GZG99). Dashed-line is the radial velocity of the stellar disk
expected from the galaxy model used in our hydrodynamic simulations
and the solid-line is that predicted for the molecular ring. Optical
emission-line velocities have been corrected from PA$\simeq$10\degr\
to PA=0\degr\ assuming a solid-body rotation for the galaxy 2D
velocity field (see GZG99). This correction is $\leq$2\% \label{fig3}}
\vspace{-3cm}
\plotone{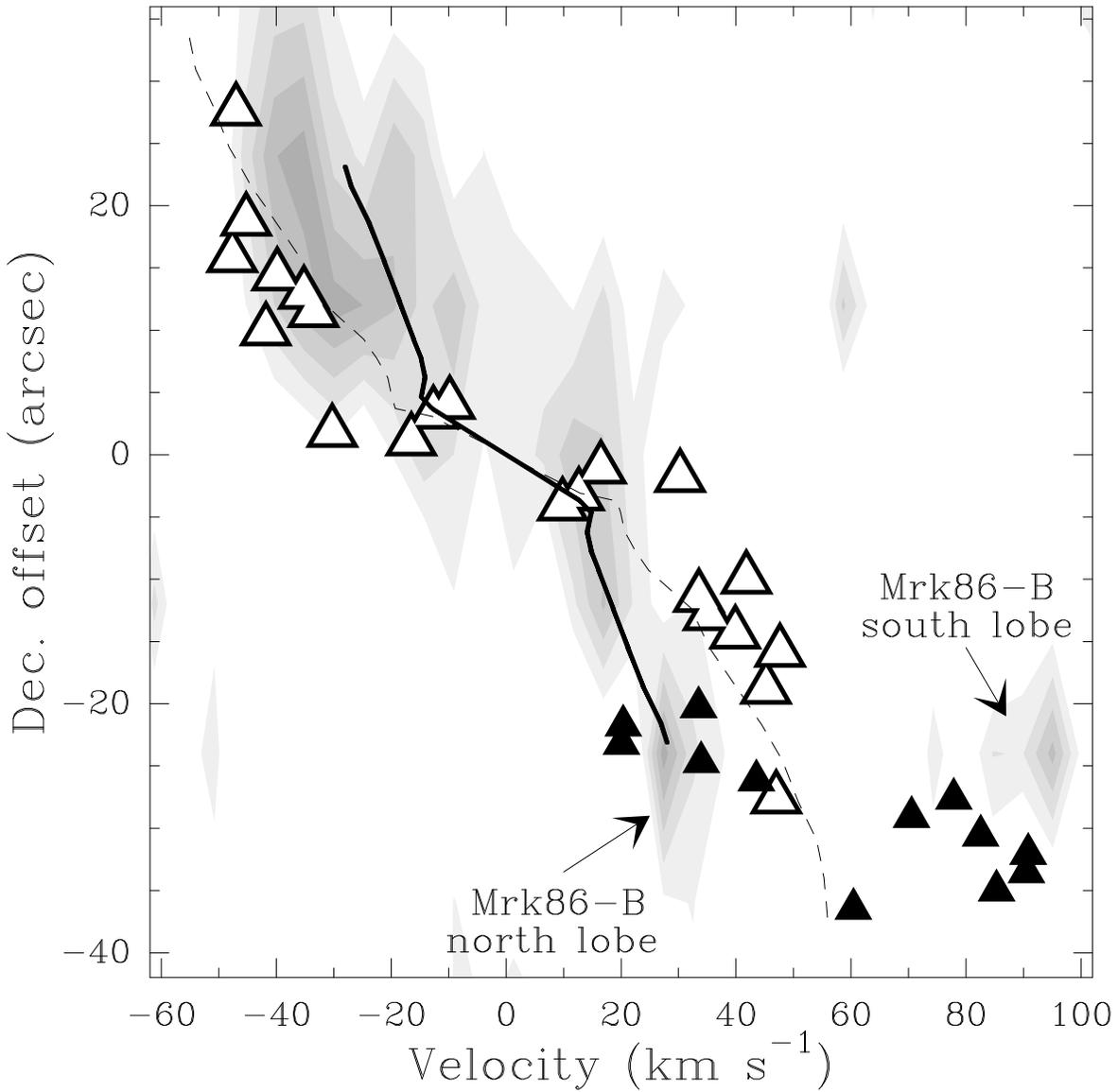}
\end{figure}

We have used a three-component model for the galaxy gravitational
field that includes stellar disk, dark matter, and an homogeneous,
spherical central star cluster. The stellar disk gravitational
potential was calculated as in Strickland \& Stevens (2000) using a
Miyamoto \& Nagai (1975) stellar-disk model with a total stellar mass
M$_{\mathrm{disk}}$=5$\times$10$^{9}$\,M$_{\odot}$ and characteristic
scale heights in the radial and vertical directions of 700 and
400\,pc, respectively. For the dark matter (DM) we used an isothermal
profile and derived the DM component parameters from the Mac Low \&
Ferrara (1999) model for a
$M_{\mathrm{ISM}}$=8.5$\times$10$^{8}$\,M$_{\odot}$ (Swaters
1999). The rotation curve derived for this model is plotted in
Figure~\ref{fig3} along with the radial velocities measured from the
ionized gas. The ISM gas distribution was derived using the Silich \&
Tenorio-Tagle (1998) semi-analytical model with a mean interstellar
turbulent velocity dispersion of 35\,km\,s$^{-1}$
derived from our optical data (GZG99).

This model indicates that the gravitational field of Mrk~86 is
dominated by the massive stellar disk component, which leads to a high
concentration of gas to the galactic plane and fast, starburst-blown,
shell blow-out into the intergalactic medium. This results in a fast
drop of the inner bubble thermal pressure, growth of the reverse shock
radius (which soon reaches a midplane shell position) and a fast
transition between the energy-dominated to the momentum-dominated
phase. Henceforth the midplane ring moves outwards under the action of
the accumulated momentum and the superwind ram pressure (Tenorio-Tagle
\& Mu{\~n}oz-Tu{\~n}on 1998). About 15-20\,Myr after the starburst
ignition the gas in the ring reaches column densities above the
critial value for molecular gas formation
($N$$>$$N_{\mathrm{crit}}$$=$5$\times$10$^{20}$\,(Z$_{\odot}$/Z);
Franco \& Cox 1986). At this moment the expansion velocity derived for
the ring is close to zero.


For a range of parameters (distance, inclination, stellar-disk and
starburst mass, gas velocity dispersion; Gil de Paz et al$.$ 2002, in
preparation) compatible with previous studies on Mrk~86 (SKT99; GZG99;
GZG00) our models lead to the formation of a ring with a H$_{2}$ mass
of about 3$\times$10$^{6}$\,M$_{\odot}$ and a radius of formation of
0.5-1\,kpc. The uncertainty in this mass can be as high as a factor of
5, considering our poor knowledge of some of the galaxy properties
required for the galaxy evolution modeling (galaxy distance, stellar
disk and starburst masses, interstellar gas metallicity, etc). This
result suggests that, if the scenario proposed by GZG00 is correct,
either the $L_{\mathrm{CO}}$-to-$M_{\mathrm{H}_2}$ conversion factor
is smaller than that predicted by the A96 relation (as it has been
discussed in the Section~\ref{sec3}) or a significant amount of
molecular gas was already present at large galactocentric distances
prior to the evolution of the nuclear starburst.


A noteworthy result of our hydrodynamical simulations is that the ring
should expand with a rotation velocity which would be below the
galactic-disk rotation velocity. A shift between the ionized and
molecular gas rotation velocities could be recognized in
Figure~\ref{fig3}. However, as we commented in Section~\ref{sec2} the
same observed effect could be partially due to beam smearing in our
CO(1-0) data. It is also worth mentioning that the difference between
the molecular ring circular velocity and the regular galactic rotation
predicted by the model can be reduced by taking into account an
exchange of mass and momentum between the stellar disk component and
the expanding molecular ring (see McKenzie et al$.$ 1978).

Finally, the observation of a molecular horseshoe instead of a
complete ring could be the result of an anisotropic evolution of the
swept out interstellar material, either due to the offset of the
nuclear starburst relative to the galaxy dynamical center
($\sim$7\arcsec\ South; GZG00) or a prior inhomogeneous distribution
of the ISM, as suggested by the asymmetry of the 21\,cm H{\sc i}-line
profile.

Summarizing, we have observed the $^{12}$CO $J$=1-0 and 2-1 emission
from the low-metallicity BCD galaxy Mrk~86. We have discovered a
horseshoe-like structure in CO emission with a total molecular gas
mass of 0.4-5$\times$10$^{7}$\,M$_{\odot}$. Both its mass and
morphology reasonably agree with the predictions of hydrodynamical
simulations for the evolution of a starburst located close to the
galaxy center. These simulations predict the accumulation of
$\sim$3$\times$10$^{6}$\,M$_{\odot}$ of molecular gas at a distance of
0.5-1\,kpc during the 20-25\,Myr of starburst-driven shell
evolution. Our numerical model also predicts a difference between the
rotation velocity of the molecular ring and the stellar disk that
could be used as an additional observational test for this scenario.

\acknowledgements We are grateful to the IRAM 30m MRT staff for their
support and hospitality. AGdP acknowledges financial support from NASA
through a Long Term Space Astrophysics grant to BFM. We also thank the
referee for his/her valuable comments, which have led to substantial
improvement of this Letter. SAS is supported by the CONACYT (Mexico)
grant 36132-E. JZ and JG are supported by the Spanish Programa
Nacional de Astronom\'{\i}a y Astrof\'{\i}sica under grant
AYA2000-1790. The research described in this paper was carried out at the Jet
Propulsion Laboratory, California Institute of Technology, under
a contract with the National Aeronautics and Space
Administration.

\newpage 

\end{document}